\documentclass[a4paper,11pt]{article}

\pdfoutput=1 
\usepackage{jinstpub} 
\usepackage{enumitem} 
\usepackage{subcaption} 
\usepackage{here}
\usepackage[numbers,sort&compress]{natbib} 
\usepackage{hyperref} 
\usepackage{lineno}
\def\beq{\begin{equation}}
\def\eeq{\end{equation}}

\def\beqa{\begin{eqnarray}}
\def\eeqa{\end{eqnarray}}

\title{\boldmath A container-based facility for testing 20'000 20-inch PMTs for JUNO}

\author[a,1]{Bj\"orn Wonsak\note{Corresponding author.},}
\author[b]{Alexander Tietzsch,}
\author[b]{Tobias Sterr,}
\author[b]{Tobias Lachenmaier,}
\author[c]{Nikolay Anfimov,}
\author[b]{David Blum,}
\author[a]{Caren Hagner,}
\author[a,b]{Benedict Kaiser,}
\author[a]{David Meyh\"ofer,}
\author[c]{Alexander Olshevskiy,}
\author[d]{Zhonghua Qin,}
\author[a]{Henning Rebber,}
\author[a,e]{Simon Reichert,}
\author[a]{Malte Stender,}
\author[f,g]{Wei Wang,}
\author[d]{Zhimin Wang}

\affiliation[a]{Institut f\"ur Experimentalphysik, Universit\"at Hamburg, Hamburg, Germany}
\affiliation[b]{Physikalisches Institut, Eberhard Karls Universit\"at T\"ubingen, T\"ubingen, Germany}
\affiliation[c]{Joint Institute for Nuclear Research, Dubna, Russia}
\affiliation[d]{Institute of High Energy Physics, Beijing, China}
\affiliation[e]{Computer Assisted Clinical Medicine, Medical Faculty Mannheim, Heidelberg University, Mannheim, Germany}
\affiliation[f]{Sun Yat-Sen University, Guangzhou, China}
\affiliation[g]{Sino-French Institute of Nuclear Engineering and Technology, Sun Yat-Sen University, Zhuhai, China}

\emailAdd{bwonsak@mail.desy.de}

\abstract{The main goal of the JUNO experiment is the determination of the neutrino mass ordering. To achieve this, an extraordinary energy resolution of at least $3\,\%$ at 1\,MeV is required for which all parts of the JUNO detector need to meet certain quality criteria. This is relevant in particular for those which are related to the energy resolution of the detector, such as the photomultiplier tubes (PMTs) to be deployed in JUNO. This paper presents the setup and performance of a dedicated PMT mass testing facility to examine and characterize the performance of the 20-inch JUNO PMTs. Its quasi-industrial size and operation level allows to test all 20000 PMTs intended to be used in the JUNO experiment. With this PMT mass testing system, several key characteristics like dark count rate, peak-to-valley ratio, photon detection efficiency, and timing resolution have been determined at an operating gain of $1\times10^7$ and assessed with respect to the requirements of JUNO. Measurement conditions and modes for the PMTs as well as estimated accuracies for the determination of the individual PMT parameters with the system are presented as well.}


\arxivnumber{2103.10193} 

\begin{document}
\maketitle
\flushbottom

\section{Introduction}
\label{sec:intro}

The Jiangmen Underground Neutrino Observatory (JUNO) experiment is a new large volume liquid-scintillator detector experiment currently under construction in the Jiangmen prefecture in Southern China \cite{junoyellowbook, junocdr}. Its main goal is to reveal the neutrino mass ordering from the neutrino oscillation spectrum of two nuclear power plants at a distance of 53\,km each, but also to contribute to a wide program in neutrino physics. 
The central detector of JUNO consists of an acrylic sphere with diameter of 35.4\,m and is filled with 20\,kt of LAB-based liquid scintillator (LS). This central detector is instrumented with about 17600 20-inch PMTs (``large'' PMTs) and 25600 3-inch PMTs (``small'' PMTs), mounted on a stainless steel truss outside of the acrylic sphere. The whole construction is embedded into a cylindrical water pool with diameter of 43\,m and instrumented with another 2400 20-inch PMTs. Thus, the water pool will be used as active Cherenkov veto against cosmic muons traversing the detector, but also acts as buffer shielding natural radioactivity from the surrounding rock. To determine the mass ordering with $3-4\,\sigma$ after six years of data taking, the detector system requires an unprecedented energy resolution of at least $3\,\%/\sqrt{E_\text{vis}\,[\text{MeV}]}$ \cite{ZhanWangCaoWen:2008, LiCaoWangZhan:2013}. 
This can be achieved only with detailed understanding of all detector components. In large volume LS detectors as JUNO, this specifically concerns all parts directly influencing the light collection and detection including the light sensitive sensors, with the large PMTs contributing about $95\,\%$ of the optical coverage of the central detector. \\
There are two types of 20-inch PMTs used in JUNO: 5000 dynode PMTs from Hamamatsu (R12860-50 HQE \cite{r12860}) and 15000 microchannel plate (MCP) PMTs from NNVT (GDB-6201 \cite{MCPRef2, MCPRef1})\footnote{The designations ``dynode'' and ``MCP'' refer to the amplification systems of the respective PMT types.}, from which 12600 NNVT PMTs and all Hamamatsu PMTs will be used to instrument JUNO's central detector\footnote{This is due to the typically better timing resolution of the selected Hamamatsu PMTs.}, while the veto will be instrumented by using only NNVT PMTs \cite{JUNO-P&D}. The performance of these PMTs crucially determines the energy resolution of the whole detector system.\footnote{The NNVT PMTs will also be preselected for a use of the best-performing PMTs in the central detector \cite{JUNO-P&D}.} Of particular concern is their photon detection efficiency (PDE) and dark noise. More parameters such as timing and pulse shape parameters are relevant for event reconstruction and background reduction. This leads to a set of requirements for the performance of 20-inch PMTs \cite{Wen:2019, Anfimov:2017}:
\begin{table}[H]
	\centering
	\caption{Performance requirements for the JUNO 20-inch PMTs. Where two values are given, they define individual requirements for the different PMT types respectively (Hamamatsu / NNVT).}
	\label{tab:pmt-reqs}
	\begin{tabular}{rcl}
		\hline
		parameter	& unit & requirement\\
		\hline
		photon detection efficiency & \% & > 24 (27 in avg.) \\
		transition time spread (TTS) for single p.e.~pulses (FWHM): & ns & < 3.5 / < 15 \\ 
		rise time, single p.e.~pulses & ns & < 8 \\
		fall time, single p.e.~pulses & ns & < 16 \\
		gain & - & 10$^7$ \\
		rate of dark noise (at $22^\circ$\,C) & kHz & < 50 \\
		peak-to-valley ratio in single p.e. spectrum & - & > 2.5 / > 2.8 \\
		pre-pulse probability & \% & < 1.5 / < 1.0 \\
		after-pulse probability & \% & < 15 / < 2 \\
		\hline
	\end{tabular}
\end{table}
\noindent
All 20000 PMTs must be tested at least once to make sure that all these requirements are met. For this purpose a dedicated multi-channel system for testing 20-inch PMTs based on commercial shipping containers has been built. The quasi-industrial size and operation level of the system is able to deal with the large number of to-be-tested PMTs and to assign all the listed PMT parameters (except the after-pulse ratio) using light distributed over the whole PMT photocathode. This allows to assess the performance of delivered PMTs for a possible use in the JUNO detector before the installation, but also to acquire detailed knowledge about the characteristics of the PMTs being important for the correct simulation and interpretation of the detector data in a very efficient way. The system is able to test both bare (using a pluggable base \cite{Qin:2018:TIPP}) and potted PMTs (with base firmly connected/glued to the PMT). \\
In this work we present all technical details of the setup and the performance of the PMT mass testing container system, and the verification of its performance with data. The test results for the full PMT sample will be presented in a separate paper. The container system is the central part within a dedicated PMT testing strategy for JUNO; this includes i.e.~a complementary system for photocathode uniformity scans (``scanning station''), see e.g.~\cite{Anfimov:2017, Butorov:2020Neutrino}. The performance of the 3-inch PMTs for the JUNO instrumentation is investigated in a different campaign \cite{sPMTs:2020}. Several other PMT test facilities were developed in the past: we mention the ones of the IMB \cite{IMB:pmttest}, Borexino \cite{Borexino:pmttest, PMTtest:Borexino}, Double Chooz \cite{doublechooz:pmttest}, Pierre-Auger \cite{Becker:2007, Barnhill:2008}, LHAASO \cite{PMTtest:LHAASO}, and KM3Net \cite{PMTtest:KM3NeT, PMTtest:KM3NeT:full} experiments. 

\section{Description of the Container System}

\subsection{Concept and setup}
\label{sec:Environment}

\begin{figure}[h]
	\centering
	\includegraphics[width=\linewidth]{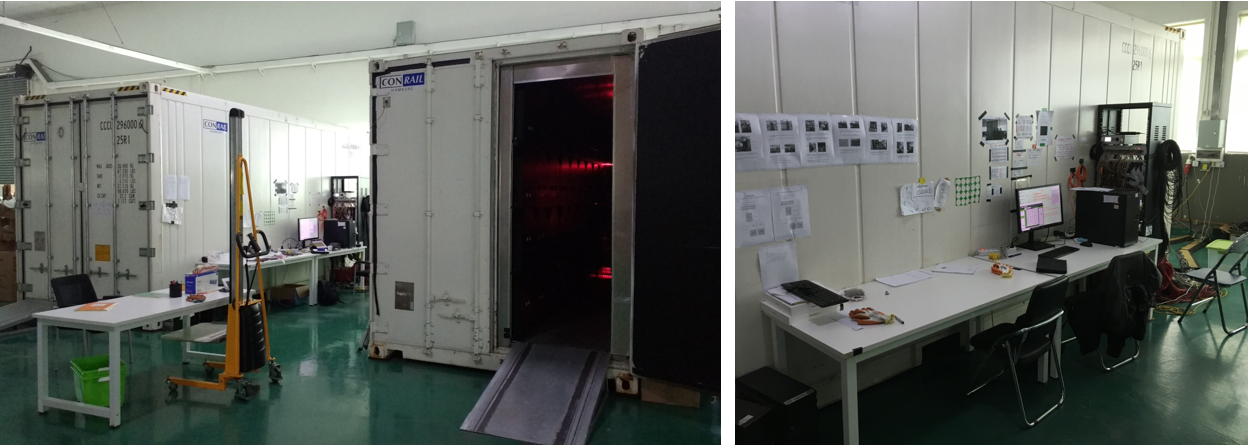}
	\caption{Pictures of the PMT container system, located in Zhongshan, China. \textit{Left:} Overview on two of the testing containers. \textit{Right:} Control desk and electronics rack.}
	\label{fig:ContainerPics}
\end{figure}
\begin{figure}[t]
	\centering
	\begin{subfigure}[c]{0.495\textwidth}
		\centering
		\includegraphics[width=\linewidth]{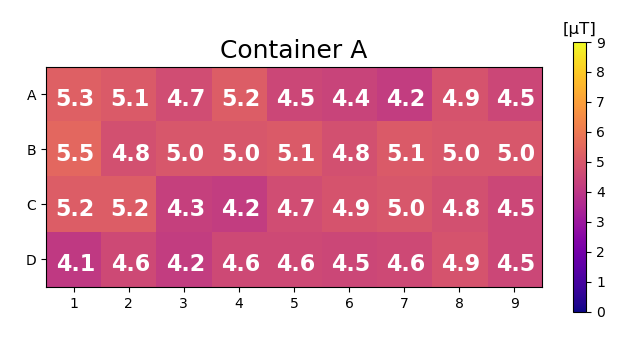}
	\end{subfigure}\hfill
	\begin{subfigure}[c]{0.495\textwidth}
		\centering
		\includegraphics[width=\linewidth]{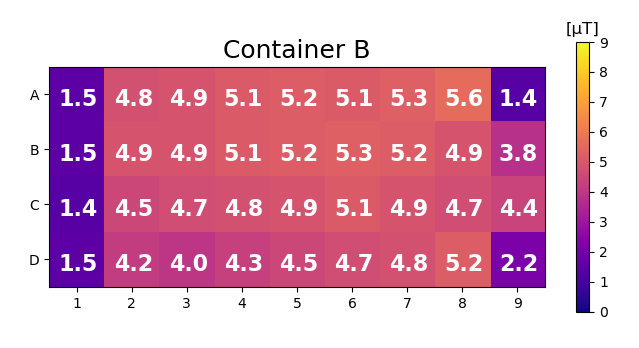}
	\end{subfigure}		
	\caption{Measurement of the residual magnetic field inside the drawer boxes of containers $A$ and $B$, performed at the PMT testing site in Zhongshan, China, at the final location and after the shipment to China. Geomagnetic field strength in Zhongshan is $45.6\,\mu$T \cite{NOAA}. The shielding of six drawer boxes in container B were reinforced by an additional layer of FINEMET (identifiable by their considerable lower residual fields).}
	\label{fig:magn-field-meas-CHN}
\end{figure}

\noindent
The basic unit of the PMT mass testing system for JUNO is a 20-feet high-cube reefer container, see figure \ref{fig:ContainerPics}. Two of these containers (denoted as ``\textit{A}'' and ``\textit{B}'') were installed at the Pan-Asia\footnote{Industrial company complex, where a conditioned storage hall (temperature about $24^\circ$\,C and humidity about $40\,\%$ over all seasons) was rented and used for PMT storage and testing.} 20-inch PMT testing and potting station in Zhongshan, China, close to the JUNO site. They act as dark rooms and provide enough space for a parallel operation of 36 PMTs per container, which are placed in optically separated measurement stations (drawer boxes, see section \ref{sec:mechanics}). The use of shipping containers thereby offers the needed versatility as the setup was mainly constructed in Germany but is operated in China.  \\
Since the performance (i.e.~PDE and local gain) of large (i.e.~the 20-inch) PMTs degrades under presence of magnetic fields as low as 10\,$\mu$T \cite{Anfimov:2017}, the interior of the containers is shielded from external fields (such as the Earth's magnetic field) by six alternating layers of silicon-iron with a total thickness of 4\,mm. The shielding material was chosen over mu-metal, because it is more robust under mechanical stress during transportation and shows an excellent shielding also against high frequency fields.
Measurements confirmed a residual field of only $4.7\,\mu$T (abs.~value)\footnote{Given value is the average over all individual measurements excluding the ones in the reinforced boxes, compare figure \ref{fig:magn-field-meas-CHN}. Measurement was performed at the final location of the containers in Zhongshan.} at the PMT positions inside the container (see figure \ref{fig:magn-field-meas-CHN}), which guarantees that influences of external magnetic fields on the PMTs are negligible. Based on experience with the shipping of the first container (container $A$), the shielding of several boxes at neuralgic points (close to door and cable feed-through) within container $B$ was reinforced with a layer of FINEMET, a nanocrystaline soft magnetic foil \cite{FINEMET}.\footnote{Magnetic shielding was observed to have been degraded during the transportation to China most likely due to mechanical shocks.}
Furthermore, the containers are equipped with a high power HVAC (heating, ventilation and air conditioning) unit, which is able to stabilize the inside temperature by $\pm1^\circ$\,C within a range from $-20^\circ$\,C to $+45^\circ$\,C \cite{thermoking:website}.
Remote control and monitoring of the HVAC unit is interfaced through a Siemens LOGO! unit. For additional monitoring, all measurement stations are equipped with temperature and humidity sensors which are periodically read out (at least every 60\,s with an accuracy of $0.25^\circ$\,C, compare \cite{Kaiser:BA}). Due to local circumstances, the HVAC unit was used only for dedicated surveys, while during the regular PMT tests we relied on the conditioning of the storage hall.\\
Container $A$ is in regular use (five working days/week) since September 2017, container $B$ since June 2018; together they provide a maximum capacity of characterizing up to 62 20-inch PMTs per day (three to five boxes per container are occupied by reference PMTs for monitoring purposes, compare section \ref{sec:calib}).  By early 2021, all 20000 20-inch PMTs have been tested at least once. 

\subsection{Drawer boxes}
\label{sec:mechanics}

\begin{figure}[h]
	\centering
	\includegraphics[width=0.8\linewidth]{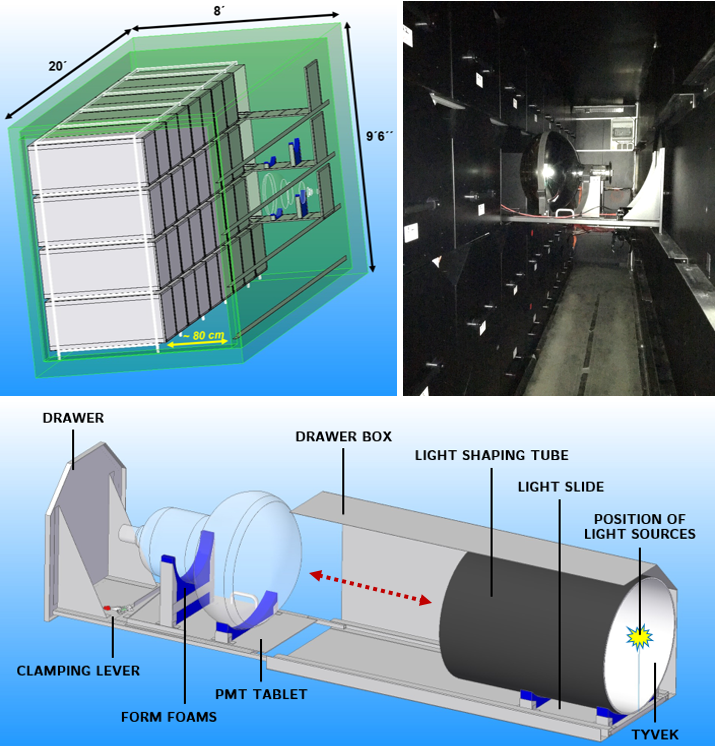}
	\caption{\textit{Top left:} Conceptual picture of one 20-feet container including the shelf system with 36 drawer boxes (in four rows). The entrance is on the front side. \textit{Top right:} Picture of the container's inside (with one PMT loaded onto a drawer), taken from the front door. \textit{Bottom:} Schematic view of a drawer box, with elements marked in the sketch (for illustrative reasons one side of the drawer box is not depicted). The light sources (not depicted here) are placed on a holder at the back of the drawer box and	aligned to the central axis of the light shaping tube.}
	\label{fig:ContainerConcept}
\end{figure}
Inside the containers, a shelf system with 36 drawer boxes (both made from aluminum) is installed, see figure \ref{fig:ContainerConcept}. The boxes are designed as small as possible to allow a maximum number of them to fit into the container. They constitute independent testing stations, each for a single 20-inch PMT. 
Further, the boxes can accommodate both PMT types (Hamamatsu and NNVT) and allow an easy loading and unloading. 
Each box features a removable drawer, two light sources (compare section \ref{sec:light-sources}), a light shaping tube made of cartridge and coated on the inside with Tyvek\textregistered~(type 1082D, $105\,\mathrm{g/m^2}$)\footnote{The Tyvek\textregistered~coverage was added to optimize the illumination of the PMT's photocathode, compare also figure \ref{fig:sim-res-final}. Tyvek\textregistered~was chosen due to its robustness and mostly diffuse reflectivity properties, compare \cite{tyvek1, tyvek2}.}, and a tray with changeable holders (form foams) to accommodate the two different types of PMTs, see also figure \ref{fig:ContainerConcept}. For the loading into the container, the PMT is fixed to the tray with two anti-static belts. The tray is then positioned on top of the drawer and fixed by a clamping lever. The mechanical parts have been produced with tolerances $< 1$\,mm, thus the PMT holders guarantee a similar position accuracy of the PMTs perpendicular to the longitudinal axis of the drawer boxes.\footnote{For the accurate positioning along this axis the PMTs are carefully aligned when they are fixed onto the tray before loading them into the drawer boxes.} Final distance between top of the PMT bulb and light sources is about 50\,cm. With some minor exceptions all mechanical parts have black surfaces (i.e.~boxes, drawers and trays). 

\subsection{Light sources and light system}
\label{sec:light-sources}

\begin{figure}[t]
	\centering
	\begin{subfigure}[c]{0.44\textwidth}
		\includegraphics[width=\linewidth]{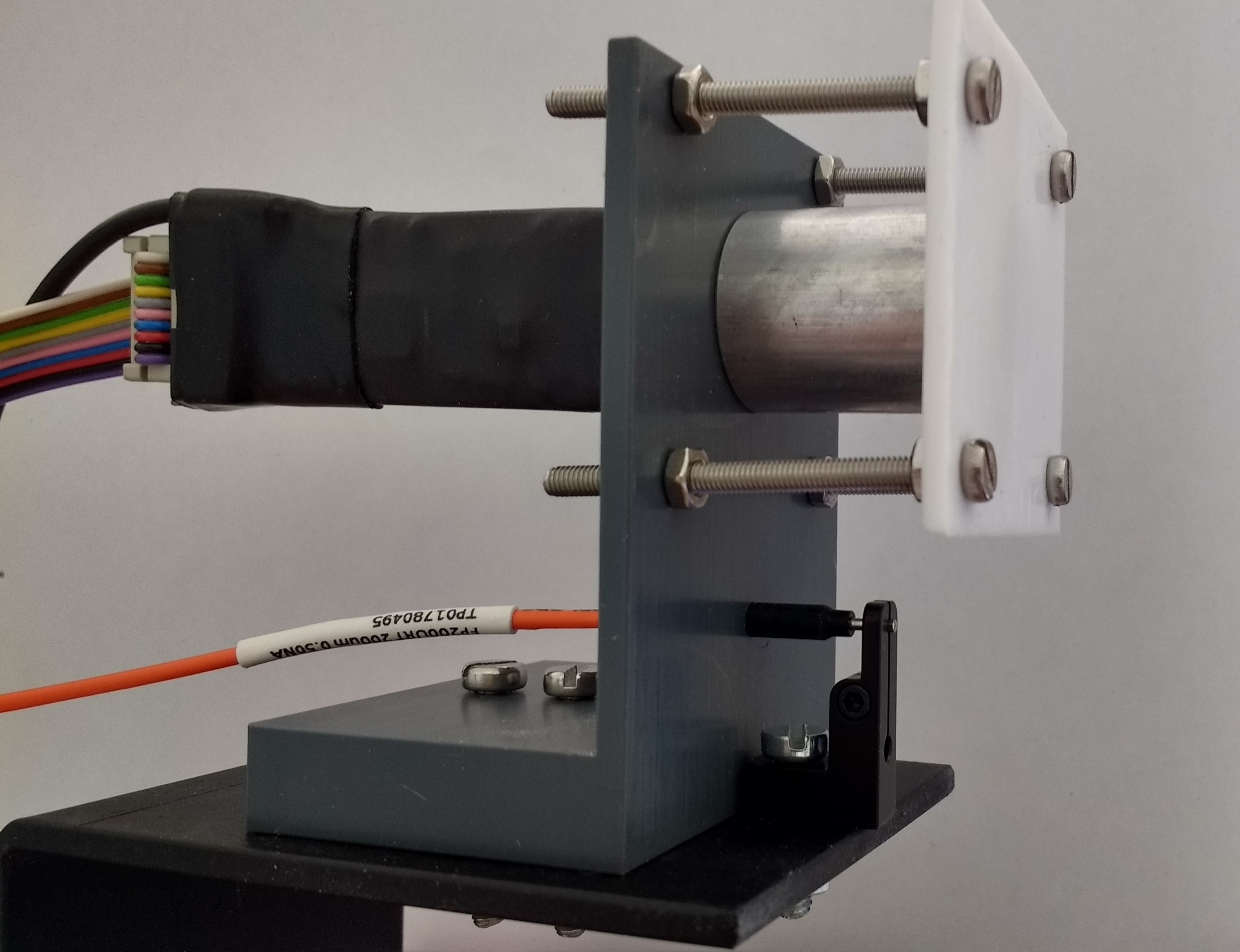}
	\end{subfigure}\hfill
	\begin{subfigure}[c]{0.54\textwidth}
		\includegraphics[width=\linewidth]{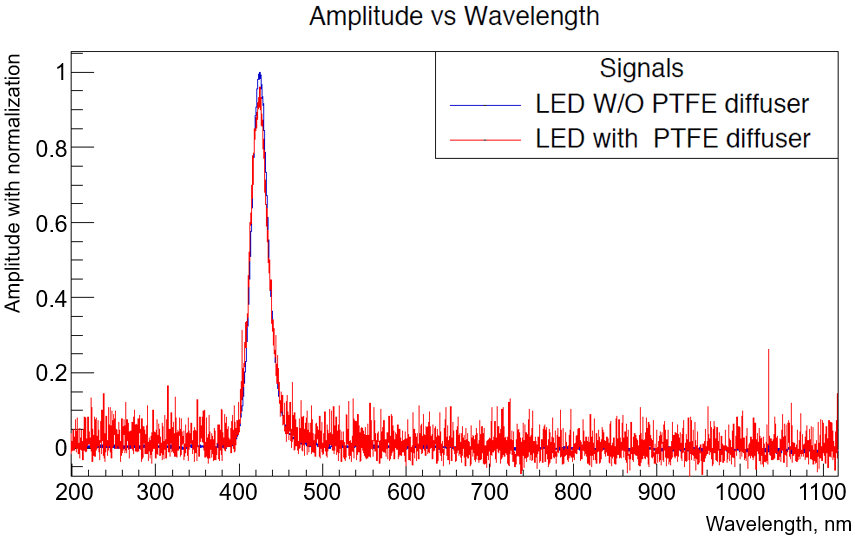}
	\end{subfigure}
	\caption{\textit{Left:} Complete light system as deployed in every drawer box. The PVC holder fixes the LED (black, with the colored flat cable attached) in position, the fiber (orange) from the laser system is also fixed by a ferrule clamp. Aluminum collimator and neutral density filter (placed inside the aluminum piece) are used for LED light attenuation. The PTFE sheet (white) in front of the LED setup is used as diffuser. \textit{Right:} Emission spectrum of the LED light pulses with and without the PTFE diffuser attached in front (normalized to maximum).}
	\label{fig:winkel}
\end{figure}

\begin{figure}[t]
	\centering
	\begin{subfigure}[c]{0.58\textwidth}
		\includegraphics[width=\linewidth]{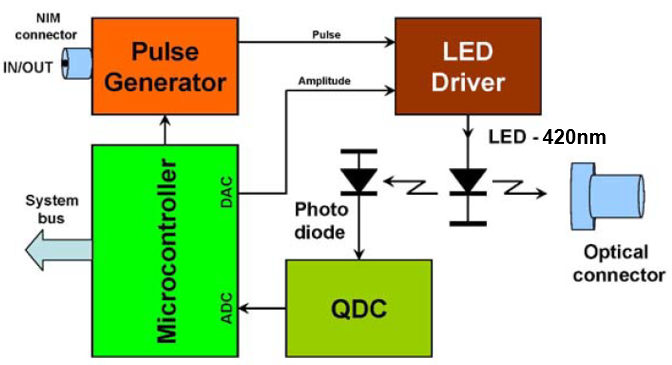}
	\end{subfigure}\hfill
	\begin{subfigure}[c]{0.41\textwidth}
		\includegraphics[width=\linewidth]{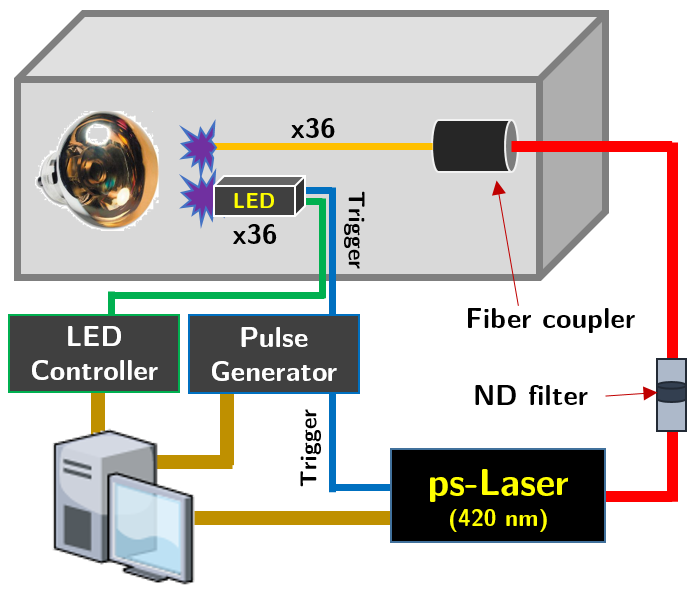}
	\end{subfigure}
	\caption{\textit{Left:} Scheme of the stabilized LED devices from \textit{HVSys} \cite{website:hvsys}, based on a STM8L151 microcontroller. For the container systems, LEDs with a wavelength of 420\,nm were used. The external trigger is provided via the IN/OUT connector by an external pulse generator. \textit{Right:} Sketch of the light distribution chain of both light systems, see \cite{Tietzsch:Diss} for more details.}
	\label{fig:led-scheme}
\end{figure}

\begin{figure}[t]
	\centering
	\includegraphics[width=\linewidth]{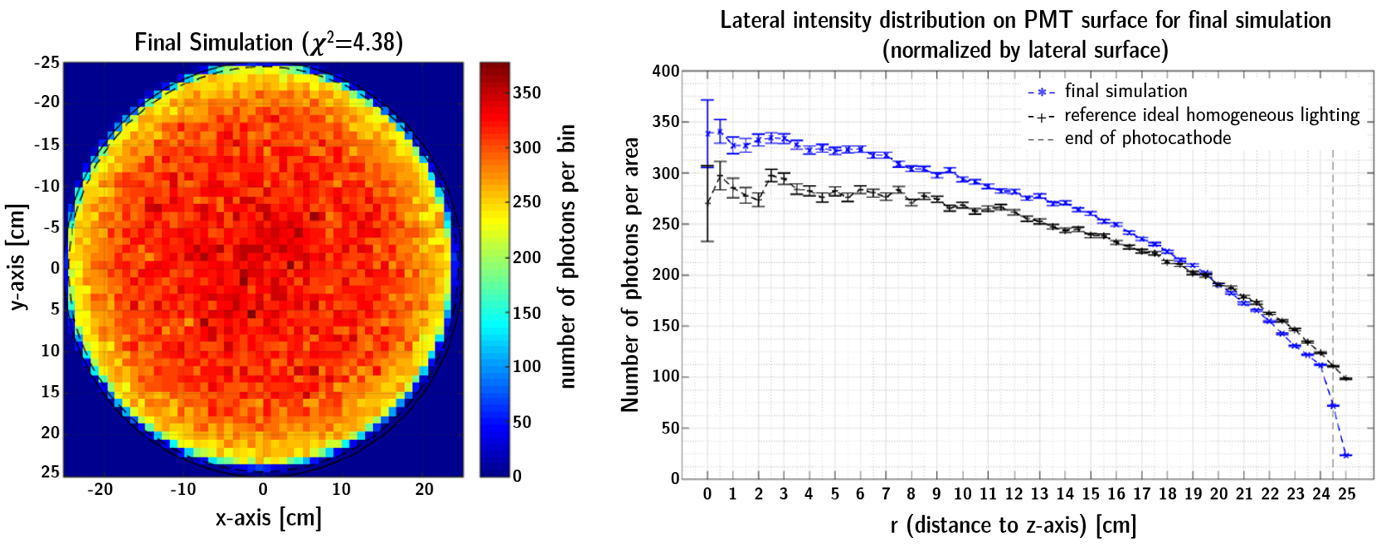}
	\caption{MC simulation result of the light field in the drawer boxes \cite{Blum:MA}, showing the photon hit distribution on the PMT's photocathode projected to the equatorial plane. Results have been confirmed in a consistency check by an independent MC simulation \cite{Reichert:BA}.} 
	\label{fig:sim-res-final}
\end{figure}

The light system of the container consists of two parts, an LED system and a picosecond laser system. They are used in different measurements (compare section \ref{sec:daq}) and can be controlled remotely. The LED system constitutes the main light source in the setup; it is used in almost all measurement steps where the PMT's photocathode gets illuminated. Only exception is the TTS measurement: here, the laser system is used due to its better timing precision. The mounting of the light sources as installed into every drawer box is shown in figure \ref{fig:winkel}. Both LED and the fiber ending from the laser system are mounted to a small PVC holder such way, that they are firmly fixed in position and their orientation is aligned to the central PMT axis. The fiber ending is additionally fixed by a ferrule clamp. The light output intensity of both systems is tailored to a Poissonian mean value level of $\mu \sim 0.05-2$ photo-electrons (p.e.) per trigger pulse. 
\\
The LED system consists of an LED device from company \textit{HVSys} \cite{website:hvsys}, which is installed into every drawer box. 
These LED devices (output wavelength of 420\,nm, monochromatic, $FWHM = 20$\,nm, see also figure \ref{fig:winkel}) deliver light pulses with a FWHM of $\sim 5$\,ns \cite{website:hvsys, Stender:MA} and are operated using an external trigger with 100\,Hz. The light intensity of the LED is regularly reading out by a PIN photodiode within the LED devices (see figure \ref{fig:led-scheme} and given reference) and used to adjust the LED power in an integrated feedback-loop. This way the light output is stabilized over time with a remaining relative amplitude variation of less than $2\,\%$ \cite{website:hvsys, Anfimov:2017INSTR}. Moreover, each LED can be addressed, monitored and the light output adjusted individually via computer, thus generating a stable and allocatable light intensity for all measurement steps and runs. The emission maximum of the LEDs in the container systems is at a wavelength of 420\,nm, which is close to the peak of the wavelength spectrum for emitted photons in the LS of JUNO. \\
The LED system has been optimized to emit only few photons per pulse: ahead of the LED device, a small aluminum collimator with 1.2\,mm diameter and a neutral density filter\footnote{The used neutral density filters feature optical depths between 2.0 to 3.0 at 420\,nm, depending on the individual brightness of the LEDs.} deployed inside are attached. In front of the collimator, a 2.0\,mm thick PTFE layer is fixed to the PVC holder and used as diffuser, see also figure \ref{fig:winkel}. The emission profile after the PTFE diffuser is very well following Lambert's cosine law\footnote{Relative decrease of $\sim 10\,\%$ at $\theta = 60^\circ$ compared to an emission profile perfectly following $\propto \cos\theta$, compare \cite{Blum:MA}.} and thus ensuring that an illumination of the complete PMT surface is achieved. This has been confirmed by measurements of the emission profile and MC simulations of the light field in the drawer boxes (considering also the effect of the light shaping tube) \cite{Blum:MA, Reichert:BA}; the expected intensity distribution on the PMT's photocathode is shown in figure \ref{fig:sim-res-final}. The emission spectrum of the light pulses however is not changed by the PTFE diffuser, see again figure \ref{fig:winkel}. \\
The second light source, a PiLas 420X (PiL042XSM) picosecond laser from Advanced Laser Systems (A.L.S.) can produce short light pulses with a wavelength of $420$\,nm and a width of $\sim 80$\,ps \cite{ALS:PILAS}, for which reason the laser system is used instead of the LEDs in the TTS measurement. 
The laser light is coupled to a fiber (multimode fiber, $200\,\mu$m core, 0.5 numerical aperture) and attenuated by an upstream neutral density filter to a level of few photons per pulse, before being guided into the container; the laser intensity is adjusted such, that a single p.e.~level is achieved for all channels. The light pulses are further distributed by a system of multiple fibers (same type as above) and guided to all boxes (compare also figure \ref{fig:led-scheme}). The splitting of the laser light into these multiple fibers happens in a small custom-made workpiece, whose design is based on \cite{Borexino:fibersys}. 
A final precision of $\leq 1$\,ns in the TTS measurement is aimed for the laser system (see also section~\ref{sec:tts}). More details about the laser system can be found in \cite{Tietzsch:Diss}. 
Light pulses from the fibers do not pass the diffuser and hit the PMT centrally on the cathode (light spot with diameter of $\sim$50\,cm is covering the full photocathode, but with large incident angles on the edges). 

\subsection{Electronics} 
\label{sec:electronics}

\begin{figure}[t]
	\centering
	\includegraphics[width=0.75\linewidth]{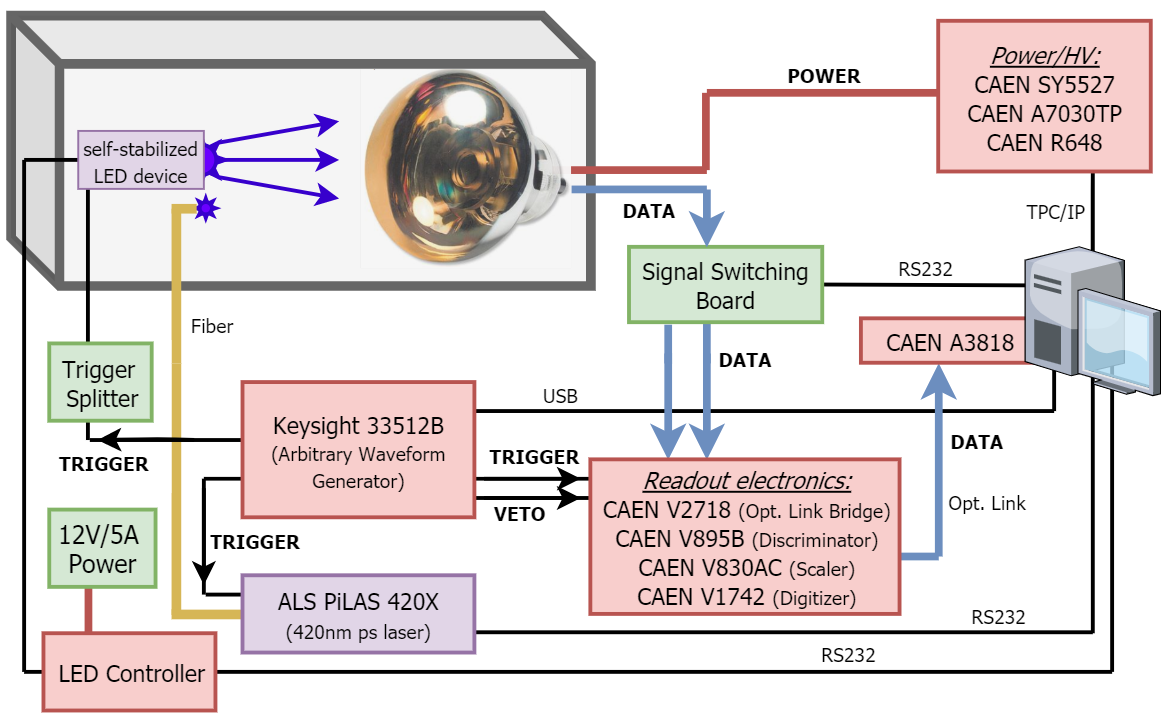}
	\caption{Electronics scheme of a single container. The electronics cover 36 channels (boxes) per container. Red boxes display commercial electronics, green boxes custom-made devices and purple boxes the light sources. All electronics devices (except the LEDs) are housed in a rack outside of the container.}
    \label{fig:elecronics_scheme}
\end{figure}

Each container is equipped with a data acquisition system covering 36 channels. A schematic overview of the system is shown in figure \ref{fig:elecronics_scheme}. All depicted electronics are controlled via PC by a fully automated data acquisition software (described in section \ref{sec:software}). \\
The HV for every channel is provided by a CAEN A7030TP board (3\,kV / 1\,mA (1.5\,W) output range
with 50\,mV resolution) in a CAEN SY5527 power supply system. 
The power supply system is connected to an interlock installed at the container door, shutting the high voltage when the door is open. This protects the PMTs from illumination with ambient light while HV is applied, but also acts as safety precaution for the operators. Two CAEN V1742 switched-capacitor digitizers\footnote{Manufacturer's calibration tables stored in EPROM on the boards are used.} per container are used for waveform acquisition. They provide 32 readout channels per module with a 12 bit resolution at $1\,\text{V}_\text{pp}$ dynamic range, combined with a sampling rate of up to 5\,GS/s and can provide a fine timing resolution of $<100$\,ps, if the triggers are recorded together with the signals \cite{V1742:manual}. 
Three CAEN V895B leading edge discriminator boards are used for counting measurements, providing 16 channels each with programmable thresholds, together with two CAEN V830AC 32-bit 32 channels latching scaler modules. The data taking electronics are read out via optical link (directly or via CAEN V2718 + A3818 optical link bridge). 
The triggers for the data acquisition and light sources are generated by a Keysight 33512B arbitrary waveform generator with a jitter (typ.~channel skew) of $< 200$\,ps \cite{Keysight:33512B}. The triggers are distributed by a custom-made trigger splitter provided by JINR.\\
The bare PMTs are operated using pluggable bases \cite{Qin:2018:TIPP}. These bases are connected to the electronics by two cables, one RG11 high voltage cable with SHV connectors (connected to the HV) and one RG58 coaxial cable (connected to the readout modules). Signals coming from the PMTs are distributed between digitizer and discriminator by a custom-made switching board (see also \cite{Tietzsch:Diss}), according on the measurement step.

\section{Container Operation and PMT Characterization}

\subsection{Data acquisition}
\label{sec:daq}

For each PMT, several parameters were measured: photon detection efficiency (PDE), dark count rate (DCR), peak-to-valley-ratio (P/V), rise time (RT), fall time (FT), single p.e.~pulse amplitude, transit time spread (TTS), prepulse ratio (PPR), and charge resolution. Before these measurements were executed, the applied HV has been set to ensure a gain of $10^7$. The afterpulse ratio of the PMTs will be determined in tests of smaller samples in the scanning station system only \cite{chen:2020:neutrino}.\\
There are four different data taking/measurement modes applied during the PMT characterization:
\begin{itemize}\setlength\itemsep{0em}
	\item $A$:	full waveform recording, using the LEDs at very low light intensity ($\mu \sim 0.1$\,p.e.), 
	\item $A'$:	full waveform recording, using the LEDs at low light intensity ($\mu \sim 1-2$\,p.e.), 
	\item $B$:	full waveform recording, using the laser system at very low light intensity,
	\item $C$:	pulse counting only, without light from the light sources 
\end{itemize}
In measurement modes $A$, $A'$ and $B$, the respective light source is pulsed with a frequency of 100\,Hz, while in mode $C$, the light sources are switched off. \\
Most of the PMT parameters investigated with the container systems can be directly extracted from the single p.e.~waveforms or its associated charge spectrum \cite{PMTbook}, except for the dark count rate.\footnote{The data analysis methods for all individual parameters will be described together with results from the PMT testing in a separate publication, expected in the second half of 2021.} All measurement, from which the actual PMT pulse form is needed for the analysis (gain, p.e.~spectrum, RT, FT, TTS, PPR, charge resolution) is recorded as waveform data, taken by the switched-capacitor digitizers (compare section \ref{sec:electronics}) with 1\,GS/s and 520\,ns windows after each trigger pulse (triggers are synchronized to the light output). Thereby, most parameters are determined in mode $A$ measurements; only the PDE is determined in mode $A'$ since a slightly higher light intensity is recommended here \cite{Anfimov:2019}. The PDE is then extracted based on the mean photo-electron (p.e.) count per trigger $\mu$ (compare section \ref{sec:calib}), which is defined as 
\begin{equation}
\mu = - \log\left(\frac{N_{q,\text{ped.}}}{N_{q,\text{tot.}}}\right) \ ,
\end{equation}
with $N_{q,\text{tot.}}$ the total number of collected charges, and $N_{q,\text{ped.}}$ the number of charges $q<0.25$\,p.e.. The TTS is determined in mode $B$ due to the better timing resolution of the laser.
The pulse counting (mode $C$, used for the DCR determination only) is performed continuously over a certain time (up to 15\,mins) with a set counting threshold of 3\,mV ($\sim 0.3\,$p.e.~at gain of $10^7$); data is taken here with the discriminator and scaler modules (compare also section \ref{sec:electronics} and \cite{V895:manual, V830:manual}).

\subsection{Measurement program and software}
\label{sec:software}

\begin{figure}[t]
	\centering
	\includegraphics[width=0.7\linewidth]{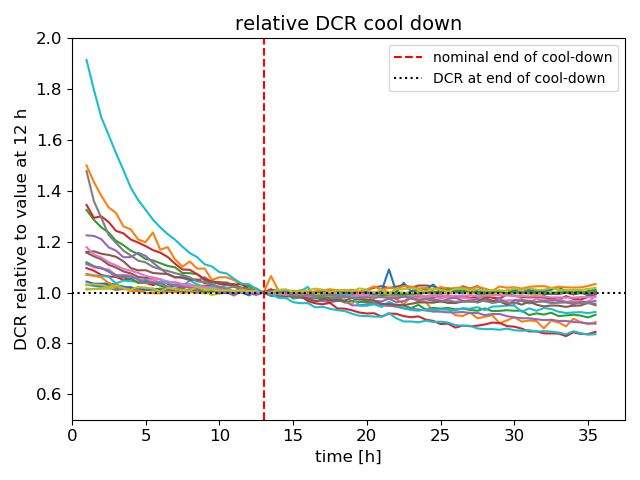}
	\caption{Development of the DCR over time for 28 PMTs (6 Hamamatsu and 22 NNVT, randomly chosen); all values normalized to values at $t=12$\,h. The PMTs have been measured for their DCR every 30\,mins after inserting them into the container. Normal behavior is observed, however not all PMTs can stabilize their rate within 12 hours (dashed line). This could lead to slightly enlarged DCR values in the PMT characterization but represents a good compromise. }
	\label{fig:Cooldown}
\end{figure}

\begin{figure}[t]
	\centering
	\includegraphics[width=\linewidth]{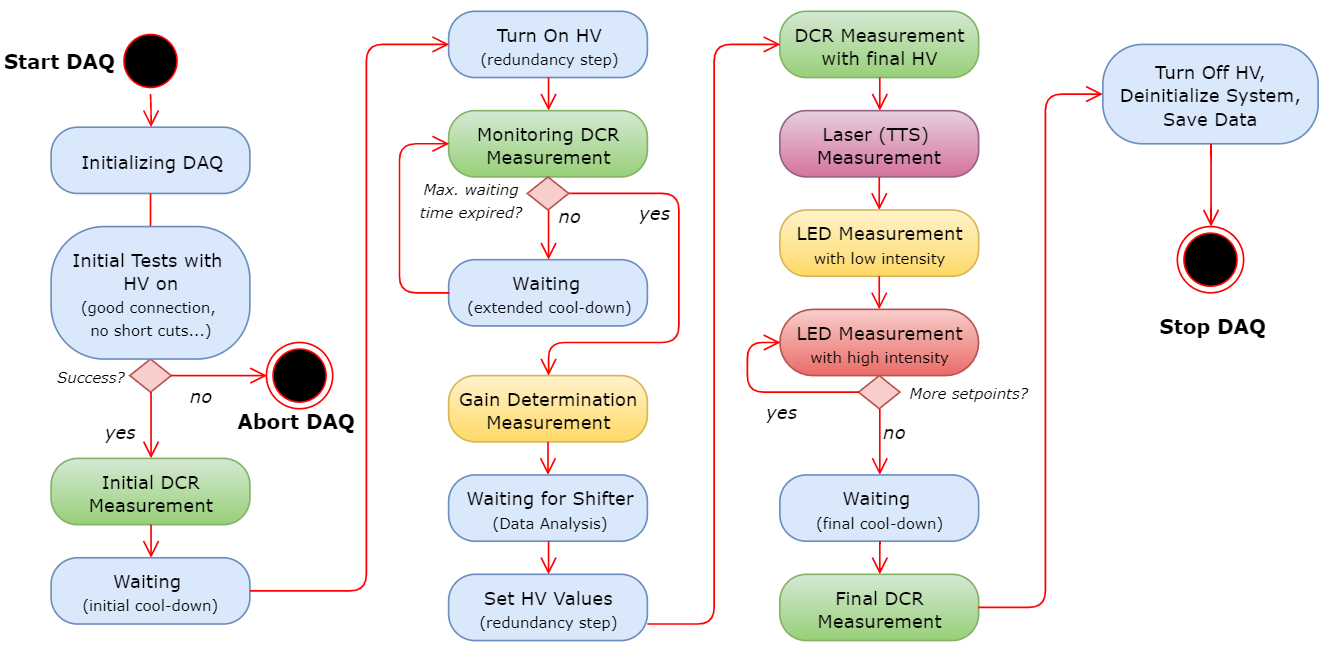}
	\caption{Complete work flow of the data acquisition software. All steps are executed automatically with only few interaction by the operator necessary. Measurement steps are indicated by colors, following the list in section \ref{sec:daq} (measurement mode $A$: yellow, $A'$: red, $B$: purple, $C$: green).}
	\label{fig:daqscheme}
\end{figure}

In order to determine all desired PMT parameters, several individual measurement steps were arranged in a sequence, which is operated for all 36 PMTs of one container load in parallel. The organization of this measurement sequence is optimized in time: the whole run fits into a 24 hours workday, including preliminary data analysis and reloading of the container. It also considers that large PMTs need some time to reduce their dark count rate after the loading into the container. 
This ``cool-down'' time was fixed to 12 hours; although not all PMTs can reach a stable DCR after this time, they however should be able to considerably reduce their dark rate in this time (see also data plot of 28 exemplary PMTs from the tests in figure \ref{fig:Cooldown}). Due to the operational constraints, 12 hours appears as a good compromise.
Further, before performing the PMT characterization, the HV has to be adjusted to ensure a gain of $10^7$ in all measurements. This is achieved by a gain/HV scan over a range of 300\,V (with steps of 50\,V) around the HV value suggested by the manufacturers for a $10^7$ gain; the measured gains at the individual steps are then fitted with a power law to determine the correct voltage. \\
The whole measurement sequence has been integrated into a LabVIEW-based data acquisition software (DAQ). This DAQ features a full automation of the whole measurement process with least remaining interactions with the operator necessary (and thus as easy as possible to control), the ability to easily access most measurement settings (easy maintenance), and a high modularity in the sequence, so that the measurement program can be easily adapted. Only few, specific information has to be provided as input, such as PMT ID (for clear identification of the recorded data), measurement channel (location of PMT inside the container) and initial supply voltage. The sequence of all individual measurement steps, as illustrated also in figure \ref{fig:daqscheme}, is operated subsequently within a total time of roughly 18 hours (including the ``cool-down''). The acquired raw data is saved locally on a server and then mirrored to servers of participating institutes (e.g.~IHEP, JINR).

\subsection{Calibration and monitoring}
\label{sec:calib}

To estimate the absolute PDE of the to-be-tested PMTs, the mean number $\mu$ of detected photo-electrons per trigger pulse is compared to a known reference value which is based on the results of 15 20-inch PMTs with known PDE, taken in a dedicated calibration campaign. In doing so, a conversion function between the p.e.~count $\mu$ for a certain LED light intensity and the actual PDE of the to-be-tested PMT was assigned to each container channel (drawer box). This also takes into account small differences between the two PMT types, compare \cite{Anfimov:2017, tietzsch:2018:neutrino}. That way, a relative measurement of the absolute PDE becomes possible with an accuracy of $1\,\%$ absolute uncertainty (compare also section \ref{sec:stab+acc}). \\
To monitor the stability and reproducibility of the measurement results from the container systems, 5 PMTs (3 Hamamatsus and 2 NNVTs) were selected and assigned to each container. These ``monitoring / reference PMTs'' are kept in the containers during all measurement runs: one of them permanently occupying the same drawer box, while the others are circulated through all measurement positions of their container. They further constitute the main instruments for the following assessment of the performance and accuracies of the containers. 
\begin{table}[t]
	\centering
	\caption{Basic information about selected reference PMTs used in containers \textit{A} and \textit{B}. First letters in serial numbers indicate manufacturers (``\texttt{EA}''/``\texttt{PA}'' for Hamamatsu/NNVT). PMTs indicated by $(\ast)$ are located at fixed position, while the others are circulating through all channels over the container runs.}
	\label{tab:refpmts}
	\begin{tabular}{cccc||cccc}
		\hline
		PMT ID & type & container & PDE\,[\%] & PMT ID & type & container & PDE\,[\%] \\
		\hline
		EA0339\,$^\ast$ & dynode & A & 26.1 & EA0574 & dynode & B & 28.0 \\
		EA0419 & dynode & A & 29.3 & EA0586\,$^\ast$ & dynode & B & 26.3 \\ 
		EA1578 & dynode & A & 24.7 & EA1437 & dynode & B & 26.5 \\
		PA1704-731 & MCP & A & 28.4 & PA1704-108 & MCP & B & 26.8 \\
		PA1705-117 & MCP & A & 28.3 & PA1705-12 & MCP & B & 26.3 \\
		\hline
	\end{tabular}
\end{table}

\section{Performance of the System}
\label{sec:results}

\subsection{Noise level and background}
\label{sec:noise}

\begin{figure}[h]
	\centering
	\begin{subfigure}[t]{0.495\textwidth}
		\centering
		\includegraphics[width=\linewidth]{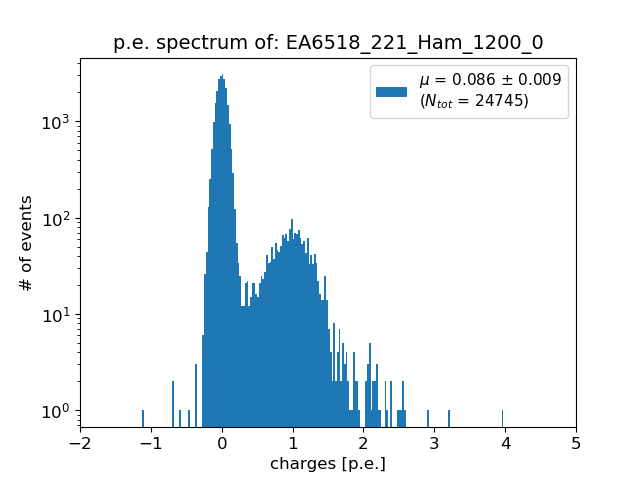}
		\caption{Single p.e.~spectrum (log-scale)}
		\label{fig:Charge_Spectra_SPE}
	\end{subfigure}
	\begin{subfigure}[t]{0.495\textwidth}
		\centering
		\includegraphics[width=\linewidth]{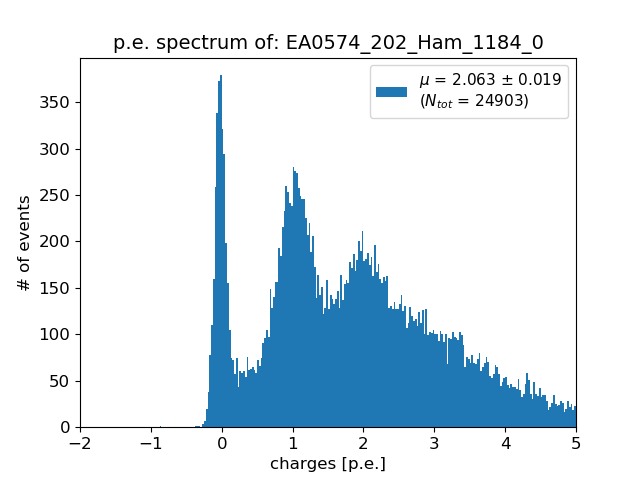}
		\caption{Multi p.e.~spectrum (linear scale)}
		\label{fig:Charge_Spectra_MPE}
	\end{subfigure}
	\caption{Single p.e.~spectrum (left, from measurement at very low light intensity ($\mu \sim 0.1$\,p.e.)) and multi p.e.~spectrum (right, from measurement at low light intensity ($\mu \sim 2$\,p.e.)) for a typical Hamamatsu PMT, tested within container $B$. The pedestal and the single p.e.~peak are clearly separated and the pedestal width is narrow. $Y$-axis is displayed in different scales for better visibility.}
	\label{fig:Charge_Spectra}
\end{figure}

Possible impact on the results from ``events'' spoiling the measurement was estimated by several independent surveys prior to the regular PMT testing period. Such events can originate e.g.~from small light-leaks in the container or optical cross-talk between boxes (background events) or from the readout electronics itself (electrical noise events). 
The light tightness of each container was checked by determining the average number of detected photo-electrons using 36 PMTs (one container load) with HV switched on in a measurement without any light sources switched on. All results are fully compatible with the individual dark count rates of the used PMTs, whereas the presence of a light-leak would have lead to an increased rate. Moreover, the results were independent from operating the PMTs with closed drawers, open drawers and even PMTs placed in the alley of the containers or when they were additionally covered with curtain, so that sufficient light tightness can be taken as guaranteed. Optical cross-talk was checked in multiple measurements with the light sources switched off in only a single channel while in all other channels they were set to max.~possible intensity. No coincident events could be observed in the respective channels and thus optical cross-talk could be excluded as well. 
Electrical noise such as baseline noise of the digitizers was estimated in similar measurements but with HV off this time. Test measurements showed a baseline noise level of $\sim 0.6$\,mV and a noise charge level of $<0.13$\,pC (corresponds to $<0.08$\,p.e.~with a gain of $10^7$) derived from the charge spectra pedestals (values averaged over all channels and consistent in both containers). The small pedestal width is visible also in the exemplary charge spectra shown in figure \ref{fig:Charge_Spectra} from regular measurements with low and very low light intensities. Noise event contributions to the counting measurements (i.e.~from the discriminator boards) have been found to be $<20$\,Hz for all channels.
\\
During the PMT testing campaign, the signal-to-noise level ($S/N$) is monitored for every PMT and every run individually based on the pedestal width $\sigma_{0\,\text{p.e.}}$ via:
\begin{equation}
S/N = \frac{Q_{1\,\text{p.e.}} - Q_{0\,\text{p.e.}}}{\sigma_{0\,\text{p.e.}}} \ .
\end{equation}
The $S/N$ ratio is observed to be stable around $12-14$ in most of the regular test runs and for all channels. In case a $S/N<10$ is found, the respective PMT will be retested in another run. 

\subsection{Comparability and stability}
\label{sec:stab+acc}

\begin{figure}[t]
	\centering
	\begin{subfigure}[t]{0.425\textwidth}
		\centering
		\includegraphics[width=\linewidth]{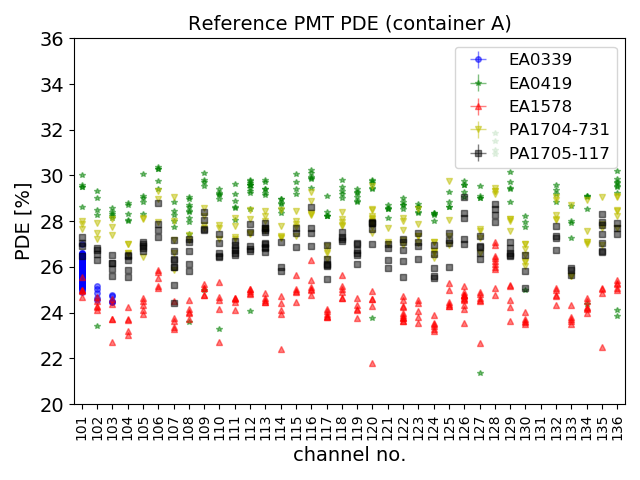}
		\caption{Container A: PDE results vs.~channels}
		\label{fig:Ref_PMTs_over_Channel_A}
	\end{subfigure}
	\begin{subfigure}[t]{0.565\textwidth}
		\centering
		\includegraphics[width=\linewidth]{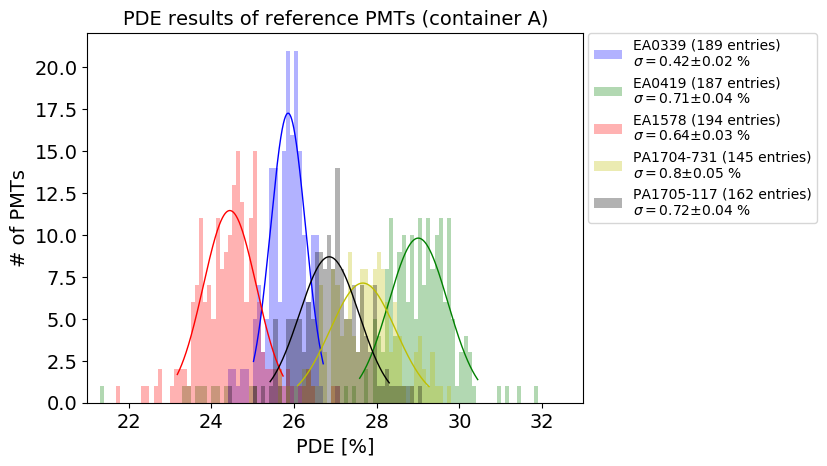}
		\caption{Container A: histogram of results separated by PMT}
		\label{fig:Accuracy_PMTs_PDE_A}
	\end{subfigure}
	\centering
	\begin{subfigure}[t]{0.43\textwidth}
		\centering
		\includegraphics[width=\linewidth]{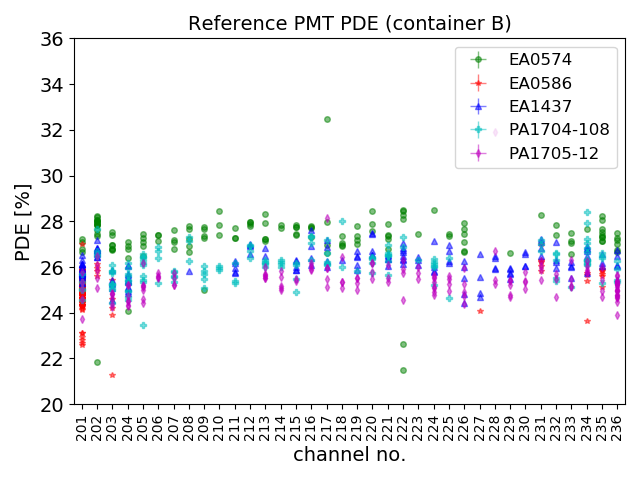}
		\caption{Container B: PDE results vs.~channels}
		\label{fig:Ref_PMTs_over_Channel_B}
	\end{subfigure}
	\begin{subfigure}[t]{0.56\textwidth}
		\centering
		\includegraphics[width=\linewidth]{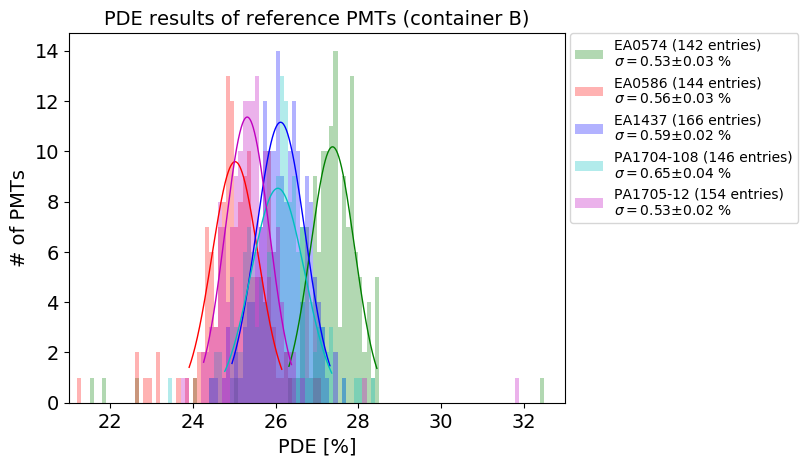}
		\caption{Container B: histogram of results separated by PMT}
		\label{fig:Accuracy_PMTs_PDE_B}
	\end{subfigure}
	\caption{PDE comparability between channels and accuracy of measurements in containers $A$ and $B$ from 200 consecutive container runs (covering data of $\sim 1$ year). The plots on the left show the measured PDE for all reference PMTs (serial numbers of Hamamatsu/NNVT PMTs start with ``\texttt{EA}''/``\texttt{PA}'') in containers $A$ and $B$ in different drawer boxes (channels). The right plots show the results of the corresponding PMTs as histograms with their width representing the accuracy of the measurements, which is less than $1\,\%$ (absolute) for all PMTs.}
	\label{fig:Ref_PMTs_over_Channel}
\end{figure}

\begin{figure}[t]
	{\centering
		\begin{subfigure}[t]{0.495\textwidth}
			\centering
			\includegraphics[width=\linewidth]{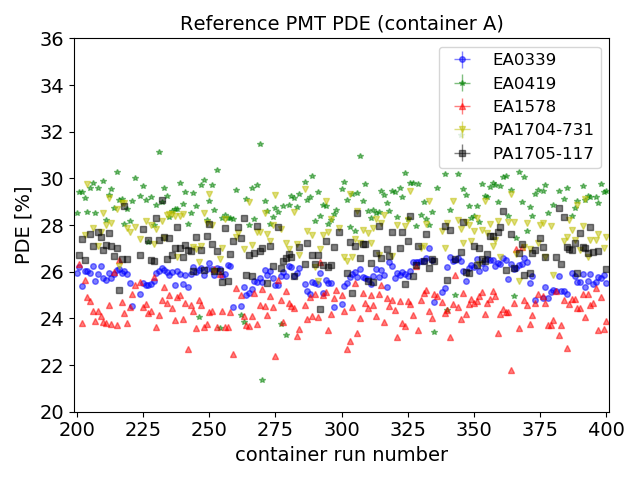}
			\caption{Container A}
			\label{fig:Ref_PMTs_over_Time_A}
		\end{subfigure}
		\begin{subfigure}[t]{0.495\textwidth}
			\centering
			\includegraphics[width=\linewidth]{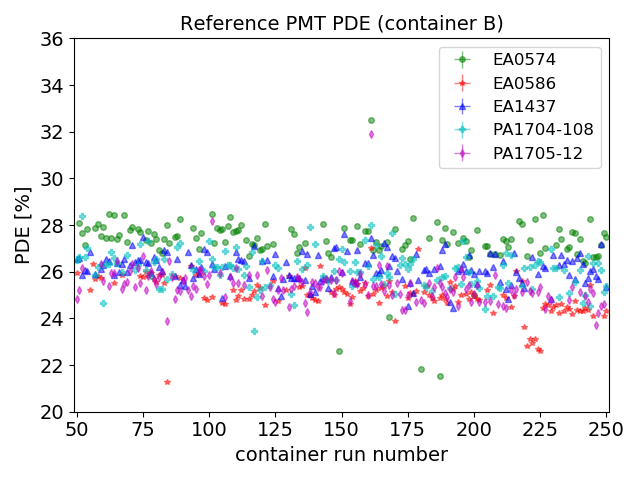}
			\caption{Container B}
			\label{fig:Ref_PMTs_over_Time_B}
		\end{subfigure}
		\caption{Measured PDE for all reference PMTs in containers $A$ and $B$ as shown in figure \ref{fig:Ref_PMTs_over_Channel}, but this time sorted by container run numbers. During normal operation, usually $5-6$ runs/week have been performed, hence the plots demonstrate the stability of the system over the time of $\sim 1$ year.}
		\label{fig:Ref_PMTs_over_Time}}
\end{figure}

\begin{figure}
	\begin{subfigure}[t]{0.33\textwidth}
		\centering
		\includegraphics[width=\linewidth]{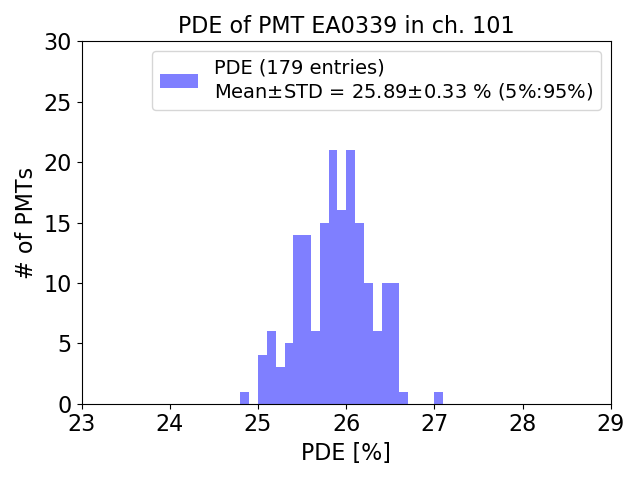}
		\caption{PDE}
		\label{fig:Ref_PMTs_same_Box_PDE_A}
	\end{subfigure}
	\begin{subfigure}[t]{0.33\textwidth}
		\centering
		\includegraphics[width=\linewidth]{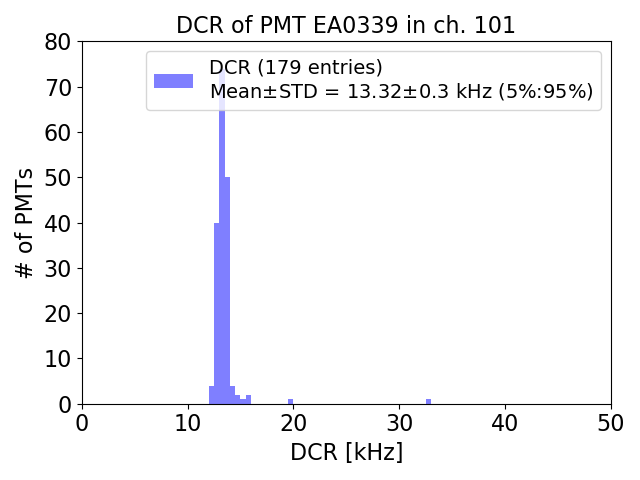}
		\caption{DCR}
		\label{fig:Ref_PMTs_same_Box_DCR_A}
	\end{subfigure}
	\begin{subfigure}[t]{0.33\textwidth}
		\centering
		\includegraphics[width=\linewidth]{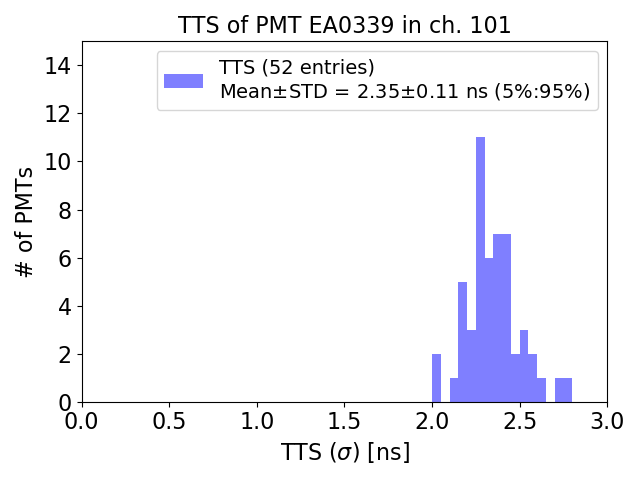}
		\caption{TTS}
		\label{fig:Ref_PMTs_same_Box_TTS_A}
	\end{subfigure}
	\begin{subfigure}[c]{0.33\textwidth}
		\centering
		\includegraphics[width=\linewidth]{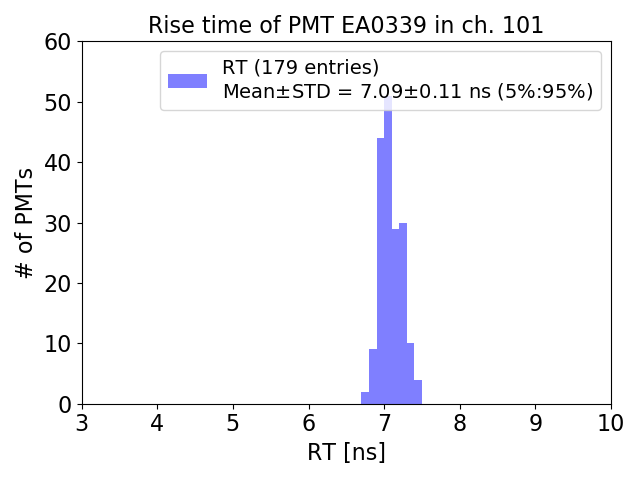}
		\caption{RT}
		\label{fig:Ref_PMTs_same_Box_RT_A}
	\end{subfigure}
	\begin{subfigure}[c]{0.33\textwidth}
		\centering
		\includegraphics[width=\linewidth]{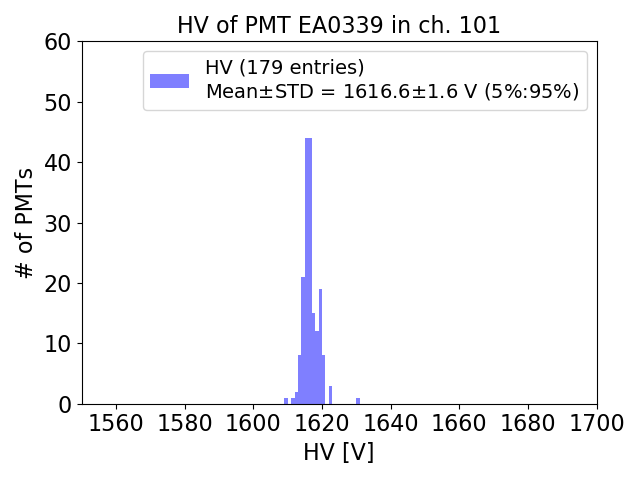}
		\caption{Appl.~HV}
		\label{fig:Ref_PMTs_same_Box_HV_A}
	\end{subfigure}
	\begin{subfigure}[c]{0.33\textwidth}
		\centering
		\includegraphics[width=\linewidth]{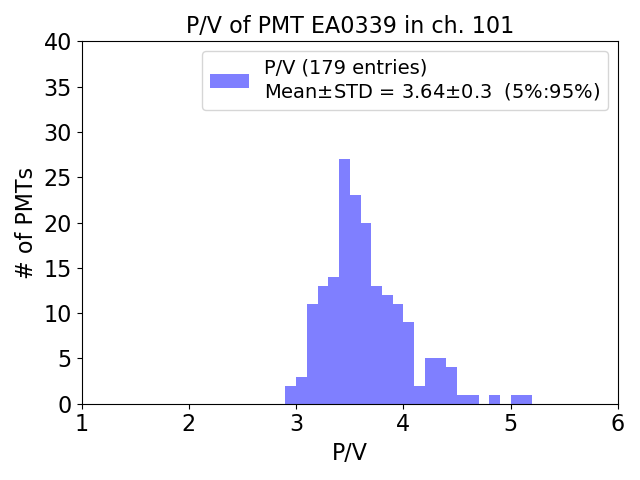}
		\caption{P/V}
		\label{fig:Ref_PMTs_same_Box_PV_A}
	\end{subfigure}
	\caption{Distribution of results for the reference PMT (Hamamatsu PMT EA0339) with fixed measurement position within container $A$ from the same container runs as above. For the calculation of the standard deviation (STD) values, only the values within the quantiles $[Q_{0.05},Q_{0.95}]$ have been used in order to exclude outliers from the sample.}
	\label{fig:Ref_PMTs_same_BoxA}
\vspace{0.3cm}
	\begin{subfigure}[t]{0.33\textwidth}
		\centering
		\includegraphics[width=\linewidth]{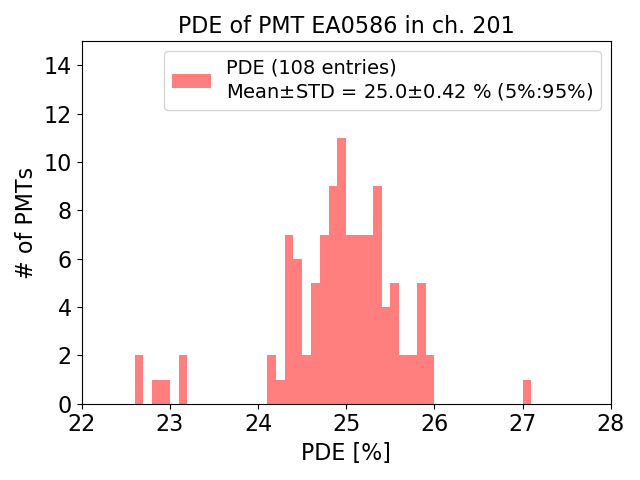}
		\caption{PDE}
		\label{fig:Ref_PMTs_same_Box_PDE_B}
	\end{subfigure}
	\begin{subfigure}[t]{0.33\textwidth}
		\centering
		\includegraphics[width=\linewidth]{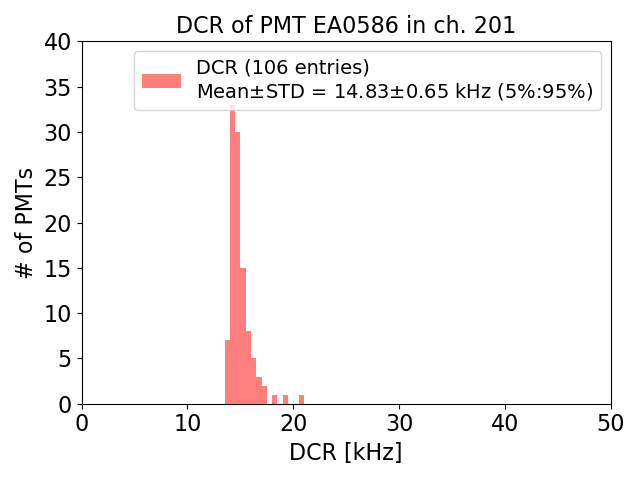}
		\caption{DCR}
		\label{fig:Ref_PMTs_same_Box_DCR_B}
	\end{subfigure}
	\begin{subfigure}[t]{0.33\textwidth}
		\centering
		\includegraphics[width=\linewidth]{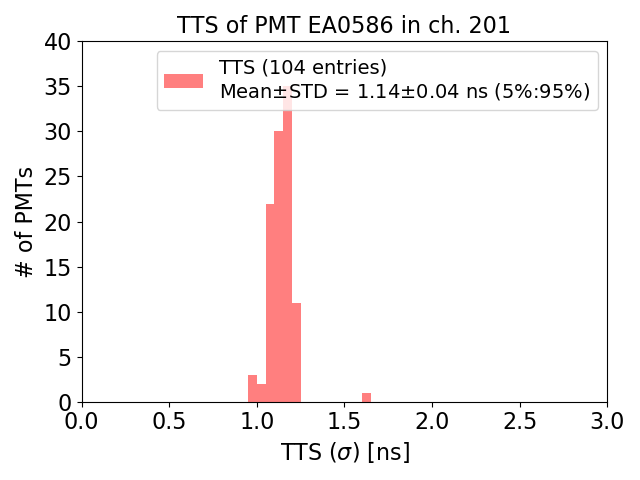}
		\caption{TTS}
		\label{fig:Ref_PMTs_same_Box_TTS_B}
	\end{subfigure}
	\begin{subfigure}[c]{0.33\textwidth}
		\centering
		\includegraphics[width=\linewidth]{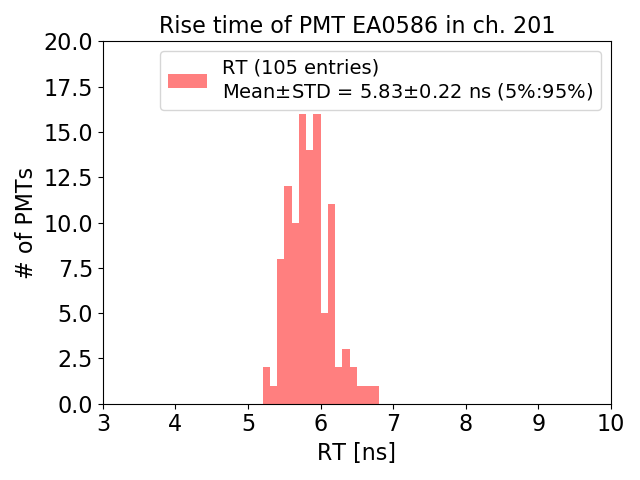}
		\caption{RT}
		\label{fig:Ref_PMTs_same_Box_RT_B}
	\end{subfigure}
	\begin{subfigure}[c]{0.33\textwidth}
		\centering
		\includegraphics[width=\linewidth]{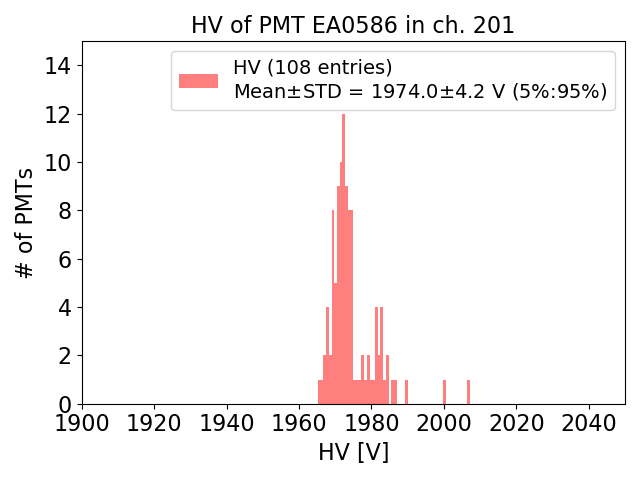}
		\caption{Appl.~HV}
		\label{fig:Ref_PMTs_same_Box_HV_B}
	\end{subfigure}
	\begin{subfigure}[c]{0.33\textwidth}
		\centering
		\includegraphics[width=\linewidth]{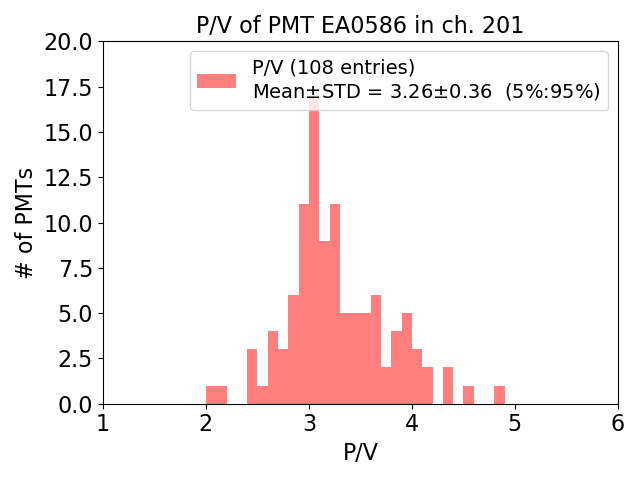}
		\caption{P/V}
		\label{fig:Ref_PMTs_same_Box_PV_B}
	\end{subfigure}
	\caption{Distribution of results for the reference PMT (Hamamatsu PMT EA0586) with fixed measurement position within container $B$ from the same container runs as above. For the calculation of the standard deviation (STD) values, only the values within the quantiles $[Q_{0.05},Q_{0.95}]$ have been used in order to exclude outliers from the sample. For the PDE measurement, a spread of less than $0.5\,\%$ (absolute) could be observed even without specifically correcting for the small downward trend visible in figure \ref{fig:Ref_PMTs_over_Time_B}.}
	\label{fig:Ref_PMTs_same_BoxB}
\end{figure}

The reliability of the measurements with the container system relies on the stability of this quasi-industrial system over the time and the comparability of the results between different channels (drawer boxes). This is verified by using the reference PMTs of each container.
Figure \ref{fig:Ref_PMTs_over_Channel} shows the measured PDE of the reference PMTs in different drawer boxes from 200 consecutive container runs, exemplarily selected to demonstrate the comparability between all boxes over the time of roughly one year. The PDE results for different drawer boxes differ within an absolute uncertainty of $<1\,\%$, which matches the design goals of the container systems and is illustrated by the result histograms on the right of figure \ref{fig:Ref_PMTs_over_Channel}. Similar analyses have been performed for several other parameters, resulting in measurement accuracies of e.g.~$<5$\,kHz for the DCR, $<10$\,V for the applied HV, $<0.4$\,ns for pulse shape parameters such as RT/FT, $<2\,\%$ for the charge resolution, $<0.3\,\%$ absolute for the prepulse ratio (PPR)\footnote{Contrary to the other listed parameters, the uncertainty of the measured prepulse ratio is mainly statistically dominated due to the relatively small absolute count numbers of prepulses (PPR expected to be $\lesssim 1\,\%$ absolute).}, and $<0.5$\,ns for the TTS, also fully matching the aimed design specifications. 
\\
\noindent
A measure for the stability of the system is shown in figure \ref{fig:Ref_PMTs_over_Time}, where the same results are plotted over the container run number and thus its development over time (here shown for the PDE, again over approx.~1 year). While in container $A$ stability of the measured data seems to be evident, a small negative trend of the measured PDE is visible e.g.~for reference PMT EA0586 of container $B$. This PMT is always kept at the same drawer (compare section \ref{sec:calib}). The different trends in both containers suggest that the effect is connected to the aging of the instrument and not the reference PMTs. Follow-up measurements have shown, that for all drawer boxes in both containers there is a negative trend in the measured PDE, but to a lesser extent ($\sim 0-1\,\%$ decrease in absolute PDE/year). These measurements indicate that the observed PDE decrease is caused by an individual behaviour of the drawer boxes. The most likely reason for this behaviour is aging of the LED devices or light distribution system. The reduction in light intensity was only observed at high light intensities (mode $A'$, compare section \ref{sec:daq}) and thus only affects the PDE measurement. It will be addressed by an additional calibration of all individual channels in both containers at the end of the PMT testing campaign and comparison with the standalone scanning station system. A contribution from aging effects of the reference PMTs themselves due to intense use however cannot be fully excluded. Nevertheless, as mentioned before, the PDE can be determined within an absolute uncertainty of less than $1\,\%$ even without an additional calibration. This can be seen in figures \ref{fig:Accuracy_PMTs_PDE_A} and \ref{fig:Accuracy_PMTs_PDE_B}, where the full aging effect is included (no compensation was applied). The resulting spread of the PDE distributions in these figures is well within the design goals of the instrument. The results for other PMT parameters show a good stability over time and comparability between different channels with no significant trend observable.
\\
\noindent
The reproducibility of the results can be checked more precisely also by focusing on the reference PMT with the fixed position (channel assignment). Figures \ref{fig:Ref_PMTs_same_BoxA} and \ref{fig:Ref_PMTs_same_BoxB} show the results for several parameters from the respective reference PMTs of containers $A$ and $B$. The PDE can be reproduced\footnote{Given values for $\sigma$ describe one standard deviation of the distribution considering only the values in the quantiles $[Q_{0.05},Q_{0.95}]$ in order to exclude outliers and thus increase the validity of the estimation (also compare the figures).} here with a spread of $\sigma_\text{PDE} \approx 0.33\,\%$ ($0.42\,\%$) absolute, while the DCR results shows a spread of $\sigma_\text{DCR} \approx 0.30\,$kHz ($0.65\,$kHz), the TTS results of $\sigma_\text{TTS} \approx 0.11\,$ns ($0.04\,$ns), the results for the risetime of $\sigma_\text{RT} \approx 0.11\,$ns ($0.22\,$ns), for the P/V ratio of $\sigma_\text{P/V} \approx 0.30$ ($0.36$), and for the to-be-applied HV of only $\sigma_\text{HV} \approx 1.6\,$V ($4.2\,$V). The PPR could be determined with a spread of $\sigma_\text{PPR} \approx 0.28\,\%$ ($0.14\,\%$). The small widths of the distributions for all parameters confirm the reliability and high reproducibility of the results for the respective drawer boxes. Both containers perform similar with slight advantage for container $A$.

\subsection{Comparison of containers \textit{A} and \textit{B}}

\begin{figure}[t]
	\centering
	\begin{subfigure}[t]{0.48\textwidth}
		\centering
		\includegraphics[width=\linewidth]{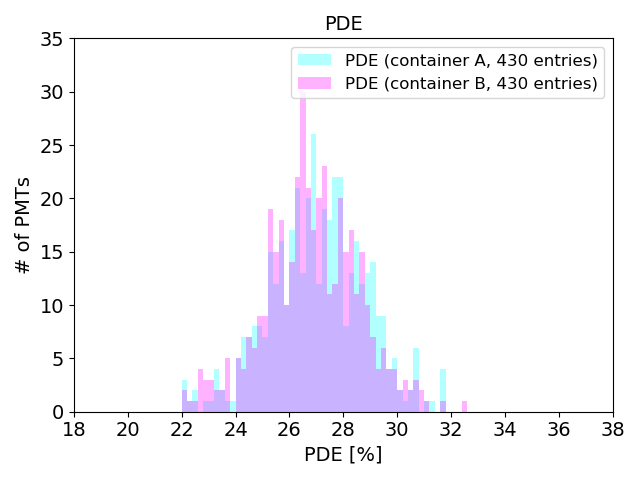}
		\caption{PDE statistics from cross-check}
		\label{fig:pde_comparison_both_container_all}
	\end{subfigure}\hfill
	\begin{subfigure}[t]{0.48\textwidth}
		\centering
		\includegraphics[width=\linewidth]{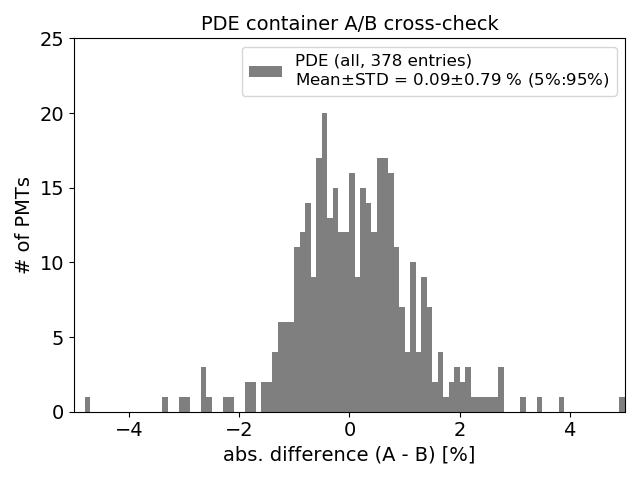}
		\caption{Absolute differences of PDE results}
		\label{fig:pde_comparison_both_container_rel}
	\end{subfigure}
	\vspace{0.3cm}
	\begin{subfigure}[t]{0.48\textwidth}
		\centering
		\includegraphics[width=\linewidth]{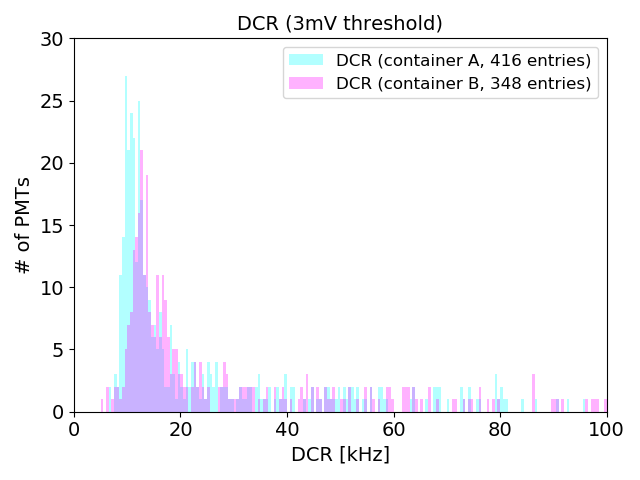}
		\caption{DCR statistics from cross-check}
		\label{fig:dcr_comparison_both_container_all}
	\end{subfigure}\hfill
	\begin{subfigure}[t]{0.48\textwidth}
		\centering
		\includegraphics[width=\linewidth]{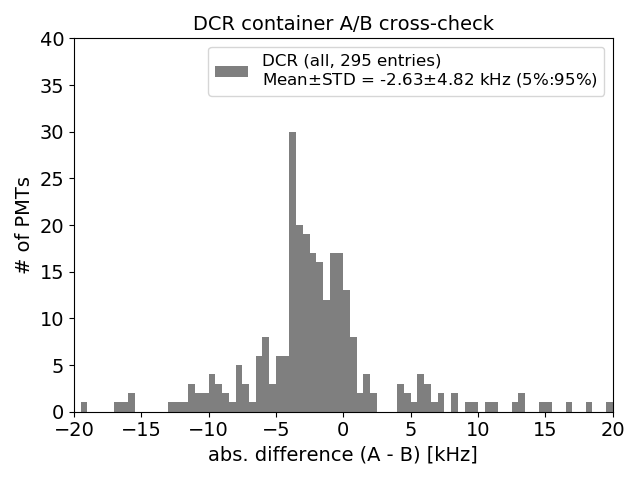}
		\caption{Absolute differences of DCR results}
		\label{fig:dcr_comparison_both_container_abs}
	\end{subfigure}
	\caption{Results for the PDE (upper plots) and the DCR (lower plots) of PMTs tested in both containers. The left plots show the statistics of the individual results, whereas the right plot shows the \textit{absolute} difference between the measured PDE/DCR values. For the calculation of the standard deviation (STD) values, only the values within the quantiles $[Q_{0.05},Q_{0.95}]$ have been used in order to exclude outliers from the sample. The observed spread for the PDE shows an absolute value of $\sim 0.8\,\%$, which corresponds to a reproducibility of $\sim 0.6\,\%$. The small bias in the distribution of differences from the DCR measurements in both containers indicates that container $B$ has an additional noise contribution of about $2.6$\,kHz in avg.~with respect to container $A$. The relatively broad width of $\sim 5$\,kHz for the DCR also contains external effects such as temperature in the storage hall and different expositions of the PMTs with ambient light during the loading. }
	\label{fig:comparison_both_container_PDE+DCR}
\end{figure}

\begin{figure}[t]
	\centering
	\begin{subfigure}[c]{0.48\textwidth}
		\centering
		\includegraphics[width=\linewidth]{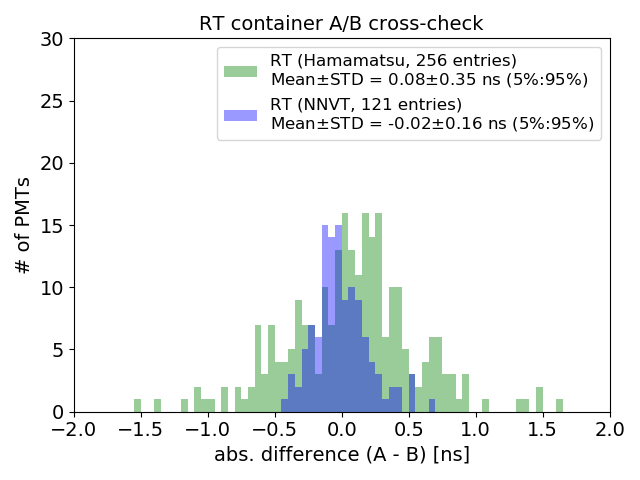}
		\caption{Absolute differences of RT results}
	\end{subfigure}\hfill
	\begin{subfigure}[c]{0.48\textwidth}
		\centering
		\includegraphics[width=\linewidth]{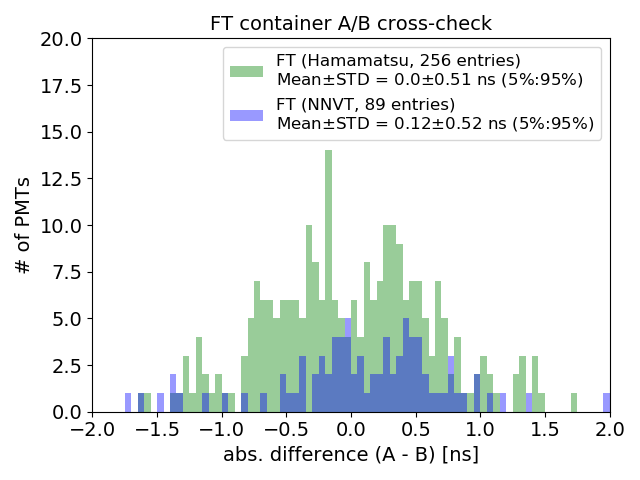}
		\caption{Absolute differences in FT results}
	\end{subfigure}\vspace{0.3cm}
	\begin{subfigure}[c]{0.48\textwidth}
		\centering
		\includegraphics[width=\linewidth]{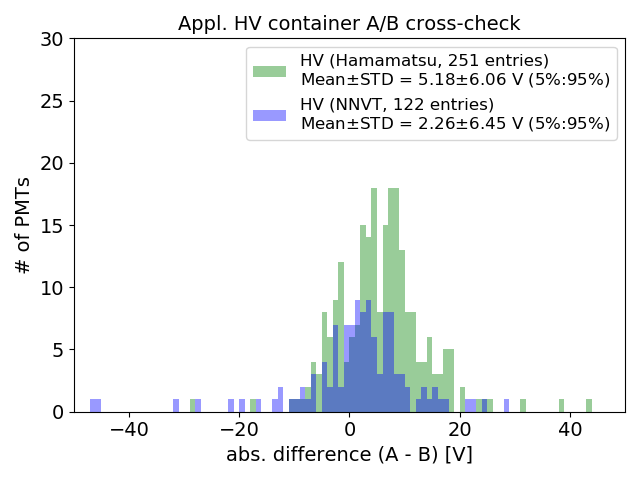}
		\caption{Absolute differences in HV results}
	\end{subfigure}\hfill
	\begin{subfigure}[c]{0.48\textwidth}
		\centering
		\includegraphics[width=\linewidth]{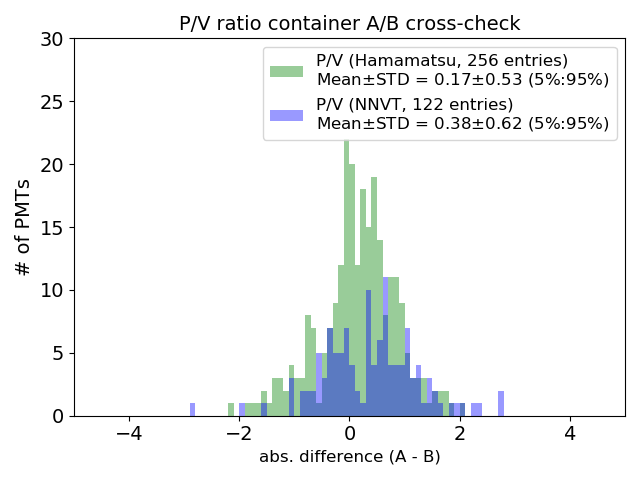}
		\caption{Absolute differences in P/V results}
	\end{subfigure}
	\caption{Difference between the results of containers $A$ and $B$ for selected parameters from a set of PMTs that have been measured in both containers. The plots show risetime, falltime, HV, and P/V results, separated by PMT type (green for Hamamatsu, blue for NNVT). The calculation of the standard deviation (STD) values was done the same way as in figure \ref{fig:comparison_both_container_PDE+DCR}.} 	
	\label{fig:more-CC-plots}
\end{figure}

As an additional cross-check, several hundred PMTs of both types have been measured in both containers. The comparison of the results provides a measure of the systematics of the individual systems. 
Figure \ref{fig:comparison_both_container_PDE+DCR} shows the results for PDE and DCR from this survey. For the PDE, the absolute difference between the measurements in both containers is $\leq 1\,\%$ indicating an absolute systematic uncertainty of the single measurement of $\approx 0.6\,\%$\footnote{Assuming that both container have the same systematic uncertainty.}, thus fulfilling the requirement of the system to measure the PDE with an absolute error of $< 1\,\%$. For the DCR, container $B$ seems to measure rates increased by $2-3\,$kHz, which indicates small differences between the set discriminator thresholds of both containers when HV is switched on.\footnote{This affects only the DCR measurement, since this is the only parameter measured using the discriminator boards. A potentially lower threshold would lead to an increased contribution of internal noise sources, though additional external (unknown) noise sources cannot be fully excluded.} 
All other parameters can be reproduced very precisely in the cross-check measurements for both PMT types, such as the to-be-applied HV for a gain of $10^7$ with $\sigma<5$\,V, the pulse shape parameters (RT/FT) with $\sigma<0.5$\,ns, and the P/V ratio with $\sigma \leq 0.5$ in average, compare also figure \ref{fig:more-CC-plots}. 

\subsection{TTS resolution}
\label{sec:tts}

\begin{figure}[htb]
	\centering
	\includegraphics[width=0.65\linewidth]{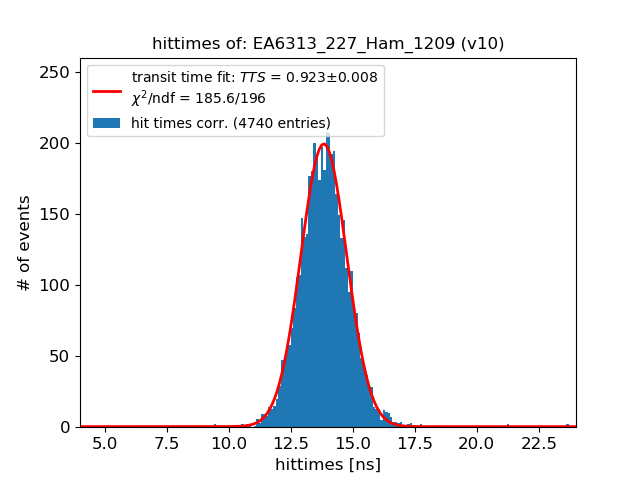}
	\caption{Hit time distribution measured with the laser system for a Hamamatsu PMT (absolute values on the $x$-axis are arbitrary). The profile shows a Gaussian distribution with a $\sigma$ of $\sim0.9$\,ns. Since the NNVT PMTs (using MCPs) have typically much larger TTS values, it is sufficient to only demonstrate the timing resolution for the Hamamatsu PMTs (dynode types).}
	\label{fig:TTS_profiles}
\end{figure}

The timing performance (respectively the resolution achievable in the TTS measurements) has been estimated by evaluating the results from the regular PMT testing of over 1000 Hamamtsu PMTs.\footnote{Since the used NNVT PMTs have much larger TTS values (typically $\sim 12$\,ns, compare also table \ref{tab:pmt-reqs}), it is more instructive to prove the container performance using Hamamatsu PMTs.}
In figure \ref{fig:TTS_profiles}, an exemplary time profile of a Hamamatsu PMT measured with the laser system in container $B$ is shown. The profile exhibits a Gaussian distribution with a width of only 0.9\,ns, which is a typical results for a well performing PMT of this type; over the whole sample, TTS values could be determined in container $B$ down to a level of $<0.8$\,ns. Therefore, an average timing resolution of about or less than $0.8$\,ns of the whole setup (container $B$) can be assumed, with similar timing resolution observed in all drawer boxes, which is sufficient to examine i.e.~the typically low TTS of the used Hamamatsu PMTs (usually $\sigma < 1.5$\,ns). 
\\
Unfortunately, the results in container $A$ do not show the same quality but reveal only a reduced TTS resolution of about $\geq2$\,ns averaged over all drawer boxes. This is most likely due to the optical fiber system in this container, which had to be refitted after some issues during laboratory tests prior to the commissioning of the containers. As a result, the light transfer is likely to be not optimal, leading to a reduced timing resolution in this container (compare also means in figures \ref{fig:Ref_PMTs_same_Box_TTS_A}, \ref{fig:Ref_PMTs_same_Box_TTS_B}). 

\section{Conclusion and Outlook}
\label{sec:summary}

We have described a quasi-industrial test facility designed to characterize 20000 20-inch PMTs for the JUNO detector. Several key characteristics of the PMTs are measured with the presented system, for which an absolute uncertainty on the PDE measurements of less than $1\,\%$ and a sufficient S/N ratio of $>10$ was achieved, as well as a good timing resolution in the TTS measurement of $<1$\,ns in container $B$. We have shown, that all PMTs are tested within adequate and stable conditions, and the system is able to produce stable and comparable results for most parameters over a large time scale and for a large number of tested PMTs. The results of both containers are in good agreement; observed systematic differences between the containers are minor and moreover quantified as reported above (slightly increased noise level in container $B$, reduced timing resolution in container $A$). At the time of this article, the PMT delivery is completed, more than 20000 20-inch PMTs have been tested with this system (including replacements for unqualified PMTs) and about $20000$ were accepted. Based on this data the PMTs eligible for JUNO will be selected and the choice of $\sim$ 17600 PMTs for its central detector will be optimized \cite{Guo:2019}. \\
Prior to the installation, the JUNO PMTs will be ``potted'' (encapsulated with the PMT base firmly glued to the PMT). To guarantee the functionality of the PMTs after potting and to cross-check their characteristics, the potted PMTs shall be tested again in the testing facility described here. To date, about 5000 of the accepted PMTs have been tested already a second time for this second functionality test. \\
Furthermore, two additional container systems with almost identical mechanics were prepared and are now in use for additional dedicated PMT tests with slightly different electronics and different purposes. One container is used to study the long-term behavior of the JUNO PMTs under normal conditions and higher stress to simulate and investigate possible aging effects. The other container was equipped with read-out electronics designed to be used in the JUNO experiment, defining a test-bed for JUNO PMTs and electronics combined. Tests with this container system will allow investigating PMTs under conditions almost identical to the ones inside JUNO. A number of follow-up papers concerning the analysis of the full PMT testing dataset, the testing of potted PMTs and the long-term behaviour of the PMTs are expected. 

\acknowledgments

This work is supported by the Deutsche Forschungsgemeinschaft (FOR2319). 

{\scriptsize \setlength{\bibsep}{0pt} \bibliography{ref/mycites_new}}
\addcontentsline{toc}{section}{References}
\bibliographystyle{JHEPmod} 

\end{document}